\documentclass[showkeys, amsmath, aps, twocolumn, superscriptaddress, prb, prstab, prstper, floatfix, 10pt
]{revtex4-2}
\usepackage{natbib} 
\usepackage{graphicx}
\usepackage{dcolumn}
\usepackage{tipa}
\usepackage{amsmath}
\usepackage[table]{xcolor}
\usepackage{xfrac}
\usepackage{hyperref}

\begin{document}

\preprint{APS/123-QED}

\title{Shape spectra of  elastic shells with surface-adsorbed semiflexible polymers}

\author{Hadiya Abdul Hameed}
\affiliation{%
UNAM - National Nanotechnology Research Center and Institute of Materials Science and Nanotechnology, Bilkent 
University, Ankara 06800, Turkiye
}
\author{Jaroslaw Paturej}
\affiliation{Institute of Physics, University of Silesia at Katowice, Chorzów 41-500, Poland}

\author{Aykut Erbas}
\email{aykut.erbas@unam.bilkent.edu.tr}
\affiliation{%
UNAM - National Nanotechnology Research Center and Institute of Materials Science and Nanotechnology, Bilkent 
University, Ankara 06800, Turkiye
}
\affiliation{Institute of Physics, University of Silesia at Katowice, Chorzów 41-500, Poland}

\date{\today}

\begin{abstract}
The shape of biological shells, such as cell nuclei, membranes, and vesicles, often deviates from a perfect sphere due to an interplay of complex interactions with a myriad of molecular structures.
In particular, semiflexible biopolymers adsorbed to the surfaces of such shells seem to affect their morphological properties. While the effect of a single, long, semiflexible chain is relatively well characterized, the mechanisms by which a high density of such surface-adsorbed polymers can alter the morphology of a spherical, soft confinement, akin to biological shells, remain relatively poorly understood. Here, we use coarse-grained molecular dynamics simulations to investigate how surface adsorption of many semiflexible polymers affects the morphology of a pressurized bead-spring shell, which is spherical in the absence of these chains. By varying the attraction strength between the chains and the shell surface, chain concentration, and the polymerization degree of chains, we demonstrate that strong surface localization of the chains can induce severe shape distortions and shrinkage, depending on the chain length and concentration. Conversely, weak localization does not induce significant shape fluctuations, yet nematically ordered phases appear on the surface. Notably, these ordered phases lead to elliptic shell shapes for chains with sizes comparable to or longer than the radius of the confinement when the elastic shell is composed of extensible, harmonic bonds. Overall, our findings offer a strategy to control the shape of synthetic shells by manipulating peripheral localization and length of semiflexible polymers while suggesting a mechanism for non-spherical shapes appearing in some biological systems.
\end{abstract}

\maketitle

\section{\label{intro} Introduction}
\begin{figure*}[ht]
\centering
  \includegraphics[width=0.9\linewidth]{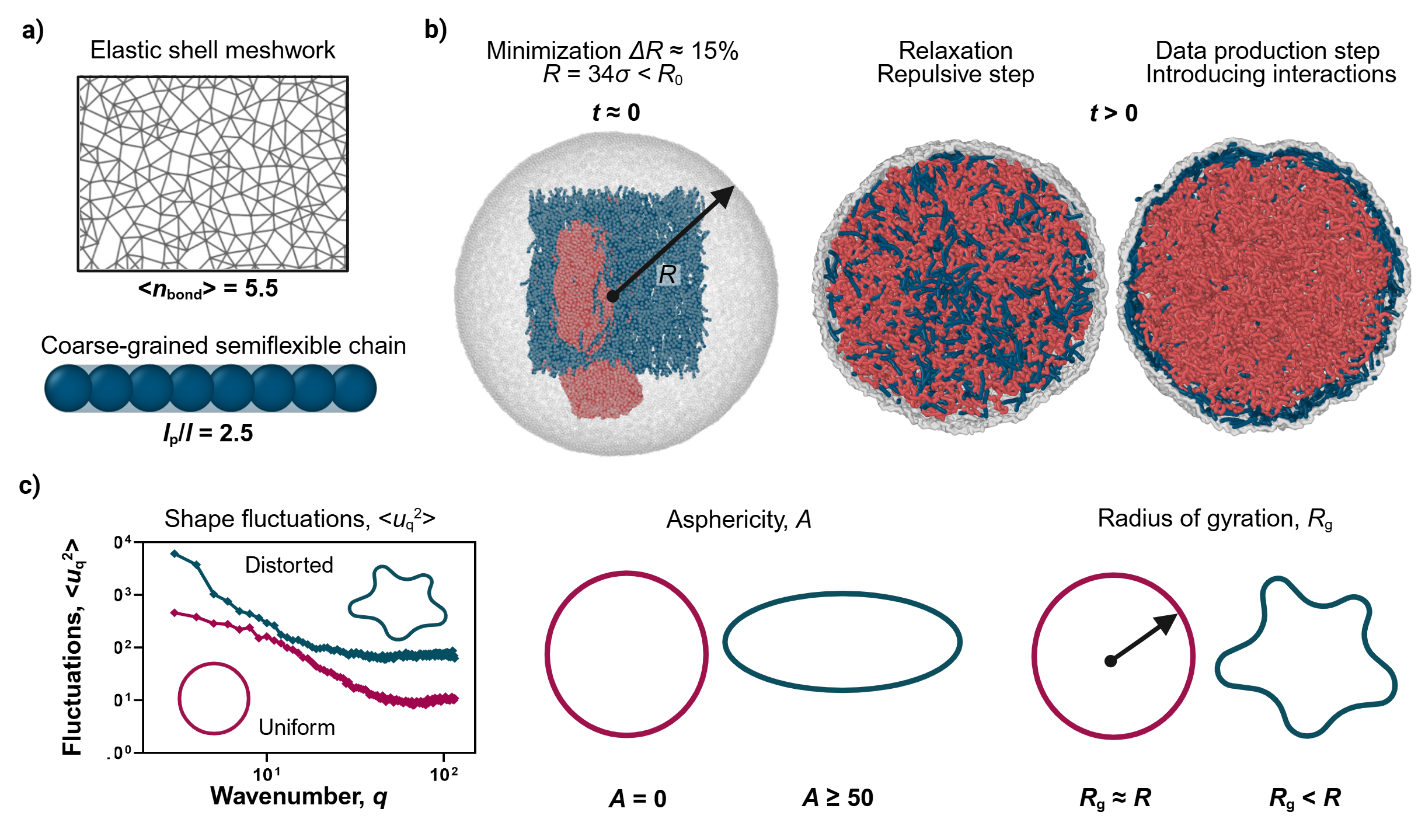}
  \caption{Schematic illustration of the coarse-grained elastic shell model, MD setup, and shape analysis metrics. a) Initial configuration of bonded elastic shell meshwork and semiflexible chains (blue).  b) 
  The initial structure is minimized to shrink the shell, then relaxed with repulsive interactions. This is followed by the introduction of attraction between rod polymers and the shell. 
  c) Time-averaged shape fluctuations, $\langle u_\mathrm{q}^2\rangle$, asphericity, $A$, and radius of gyration of the shell, $R_\mathrm{g}$, are used to quantify shell shape anomalies.
  }
  \label{fig1}
\end{figure*}
The shape regulation of biological shells plays a fundamental role in the stability and functionality of various biological systems. For instance, the force interplay between the viscoelastic actin network making up the cytoplasm and the deformable cell membrane dictates cell shape and size~\cite{Dmitrieff2017, Koenderink2015}. Similarly, cellular compartments such as the nucleus experience shape alterations due to chromatin pressure at the interior~\cite{Attar2024},  the mechanics of the cytoskeleton, and nuclear lamina, a layer of filamentous proteins lining the inner nuclear membrane~\cite{Bausch2016, Nir2019, Cohen2004, Turgay2017, Hameed2024}. Aberrant nuclear shapes are considered hallmarks of several diseases, such as progeria, muscular dystrophy, and even certain types of cancer~\cite{Flores2021, Denais2016, Irianto2016, Butin-Israeli2012, Misteli2010}, suggesting the importance of shape regulation mechanisms in those systems. 

Many biopolymers that could intervene in the shape of the above-mentioned \textit{shells}, such as double-stranded DNA, actin, or lamin supramolecular structures,
appear to be of a semiflexible nature.
Semiflexibility arises if the persistence length, $l_\mathrm{p}$, of the polymer is comparable to its contour length, $l$~\cite{Rubinstein2003}.  
The persistence length, $l_\mathrm{p}$, defines the characteristic length scale below which thermal fluctuations cannot bend the polymer chain~\cite{Rubinstein2003, Gutjahr2006, Girard2024, Kierfeld2008}. 
On length scales shorter than or comparable to $l_\mathrm{p}$ (i.e., $l_\mathrm{p} \geq l$), the polymer can behave as a rod-like or semiflexible polymer, respectively~\cite{Rubinstein2003, Gutjahr2006, Girard2024, Kierfeld2008}.
The persistence length scales with the thickness of a polymer, $a$, as $l_\mathrm{p} \sim a^4$, which renders many biological fibers as highly stiff, rod-like chains at short length scales. For instance, double-stranded DNA with a thickness of $a=2$ nm has a uniform persistence length of $l_\mathrm{p} \approx 50$ nm, which constitutes $\gg 150$ base pairs, elucidating its semiflexible nature~\cite{Marko1994, Kampmann2013, Girard2024, Bednar1995}. Moreover, lamin protofilaments with $a \approx 4$ nm could exhibit persistent length values at around  $l_\mathrm{p} \approx 500$ nm~\cite{Turgay2017}.

If such semiflexible biopolymers are confined inside a deformable shell, such as a lipid vesicle or a hollow-hydrogel shell, they can compete with the bending rigidity of the confinement and lead to various shape fluctuations of the shell;
%
if a single semiflexible polymer (e.g., actin) has a contour length, $l$, larger than the vesicle radius, $R$, the buckling of the chain is accompanied by increasingly aspherical vesicle shapes~\cite{Li2018, Shi2023, May2013}.
%
Notably, increasing confinement by increasing the persistence length (or decreasing the confinement size) localizes the chains towards the periphery in order to minimize chain bending. In turn, this local concentration increase enhances nematic ordering~\cite{Girard2024, Gao2014, Marenduzzo2007, Spakowitz2003}, further contributing to the shape distortion of shells~\cite{Li2018, Guo2009, Li2025, Shi2023, May2013}.

When ``many'' such semiflexible chains or their supramolecular structures are confined inside a soft shell, the shape distortion often requires chain adsorption to the confining walls~\cite{Bausch2014}.
In fact, if the confinement is weak, which can be controlled by decreasing chain concentration or the chain contour length, $l$, relative to the vesicle radius, $R$~\cite{Peterson2021, Bausch2014}, the adsorbed chains can form nematic-phase domains or ``tennis ball''-like patterns on the surface of the confinement~\cite{Nikoubashman2017, Binder2018, Zhang2011, Hameed2024}. Nonetheless, adsorbed chains do not change the spherical form of the vesicle at weak confinement~\cite{Bausch2014, Bausch2016, Erwin2010, Liu2008}. Under strong confinement (i.e., increasing chain concentration or contour length, $l$), the chains can form global phases where most chains align in a parallel fashion on the vesicle surface~\cite{Milchev2017, Khadilkar2018, Binder2018, Nikoubashman2021}, which, in turn, can also generate elongated or non-spherical vesicle shapes~\cite{Bausch2014, Liu2024}. 
Consistently, implicit modeling of pressurized elastic shells showed that the shape can transition from spherical to an ellipsoid if the shell is composed of anisotropic structural elements~\cite{Maji2023}. 
%


In the context of biology, structures such as the nuclear lamina meshwork are adsorbed to the confinement walls (i.e., nuclear boundary) and can govern the shape of the cell nucleus~\cite{Turgay2017, Nmezi2019, Attar2024, Wiesel2008, Stephens2017}. Notably, in several disease types, lamin structures reminiscent of nematic-phase domains appear on the nuclear surface~\cite{Scaffidi2006, Hameed2024, Misteli2010, McCord2022, Dahl2006}. Similarly, dsDNA scaffolds or DNA origami structures mechanically strengthen vesicles~\cite{Okano2018, Fu2017, Bhatia2020}, but distort the shape at high concentrations~\cite{Schwille2015}. 
Overall, while the effect of chain concentration on the phase behavior of such semiflexible polymers or active filaments has been discussed~\cite{Hameed2024, Peterson2021, Naturephy2020, Naturephy2025}, how this self-assembly on the surface and parameters that can control surface-adsorption could affect the shape of an elastic shell, remains relatively poorly understood. 

In this study, we use coarse-grained molecular dynamics (MD) simulations to study how the adsorption of semiflexible polymers to the internal walls of a pressurized elastic shell could change its shape. The elastic shell was developed in our previous studies and can change its form from a perfect sphere~\cite{Attar2024, Attar2025}. Motivated by lamin-protein-driven nuclear shape anomalies in several diseases, we consider chain densities below and above the chain overlap concentration, alongside chain lengths below and above the dimensions of the confinement, to emulate various confinement scenarios.


Our results show that adsorption of polymers to the surface can cause a wide spectrum of shape fluctuations from spherical forms with large undulations to elongated morphologies. Softer shells are more prone to form oval structures by nematically oriented chains on the surface, whereas strong adsorption distorts these phases and leads to shape fluctuation in a wide range of wavelengths. Our calculations demonstrate the effectiveness of combining polymer and vesicle elasticity to design hybrid structures, while qualitatively addressing nuclear shape anomalies observed in several genetic diseases~\cite{Flores2021, Denais2016, Irianto2016, Butin-Israeli2012, Misteli2010}.

\section{\label{methods}Methods}
%
\subsection{\label{m1}Simulation model}
Our MD simulation model emulates a pressurized elastic shell, where a semiflexible polymer can adsorb onto the confining interior walls of a soft, deformable shell. The model consists of three components:  i) flexible polymers providing an internal pressure, ii)  semiflexible chains that can adsorb to the surface,  and iii) an elastic shell confining all the polymers (see Fig.~\ref{fig1}a). 
All components are represented using a coarse-grained approach, where each constituent is modeled as a collection of spherical beads (monomers) of mass, $m$. These monomers are connected either into chains (for flexible polymers and semiflexible chains) or into a mesh-like network (for the elastic shell) via springs to capture the mechanical connectivity (see Fig.~\ref{fig1}a). All polymer chains consist of monomers linked sequentially, differing in their bending stiffness, while the shell is modeled as a monolayer meshwork of monomers connected to their nearest neighbors.

Non-bonded interactions between all monomers are described by a shifted and truncated Lennard-Jones (LJ) potential
%
\begin{equation}
    U_\mathrm{LJ}(r_\mathrm{ij}) =
    \begin{cases} 
        4 \epsilon \left[ \left( \frac{\sigma}{r_\mathrm{ij}} \right)^{\alpha} - \left( \frac{\sigma}{r_\mathrm{ij}} \right)^{6} \right] & \text{for} \quad r_\mathrm{ij} < r_\mathrm{c}, \\
        0, & \text{for} \quad r_\mathrm{ij} \geq r_\mathrm{c}.
    \end{cases}
    \label{eq:lj}
\end{equation}
where  $r_\mathrm{ij}$ is the distance between two interacting beads $i$ and $j$, $\epsilon$ is the interaction strength, $\sigma$ is the bead diameter and $\alpha$ specifies the exponent of the repulsive term. 
In this study, we use reduced LJ units by setting
$\epsilon = 1$ and $\sigma = 1$, so that energy and length are expressed in these units. 
The corresponding units of temperature and time are: $[T] = \epsilon/k_\mathrm{B}$,  and $[\tau] = (m\sigma^{2}/\epsilon)^{1/2}$, respectively. 
The cutoff distance for monomer interactions is set to $r_\mathrm{c}=2^{1/6}\sigma$ and exponent $\alpha=9$.  
The same parameters are used for repulsive interactions between flexible polymers and the shell. In contrast, attractive interactions between the semiflexible chains and the shell, $U_\mathrm{RS}$, are introduced by using a longer cutoff  $r_\mathrm{c}=2.5\sigma$ and larger exponent $\alpha=12$. 
%
By varying the exponent of the repulsive term $\alpha$ (e.g., from 9 to 12), we effectively modify the steepness of the short-range repulsive interaction. A higher exponent leads to a steeper and more rigid repulsive interaction, whereas a lower exponent produces a softer repulsion and a shallower potential minimum (for the same value of $\epsilon$)~\cite{HansenMcDonald2013, Likos2001, Smith2003}.

The bonded interactions between monomers are modeled by a Finitely Extensible Non-linear Elastic (FENE) potential unless noted otherwise
\begin{equation}
   U_{\text{Bond}}(r_\mathrm{ij}) = -0.5 k r^2_\mathrm{ij} \ln \left[ 1 - \left( \frac{r_\mathrm{ij}}{r_\mathrm{0}} \right)^2 \right] \quad \text{for} \quad r_\mathrm{ij} < r_\mathrm{0}.
   \label{eq:fene}
\end{equation}
where $k$ is the spring constant, and $r_\mathrm{0}=1.5\sigma$ is the maximum bond extension~\cite{Kremer1990}.
 We use bond stiffness $k=30.0 \mathit{k}_{\mathrm{B}}\mathit{T} / \sigma^2$ for flexible polymers and semiflexible chains. For bonded shell monomers, we apply lower bond stiffness, $k = 5.0 \mathit{k}_{\mathrm{B}}\mathit{T} / \sigma^2$, making the shell bonds more prone to deformation~\cite{Kremer1990, Nikoubashman2021}.

The bending rigidity of semiflexible polymers is governed by a harmonic bending potential
%
 \begin{equation}
     U_{\text{Bend}}(\theta) = k_{\theta} (\theta_\mathrm{ijk} - \theta_\mathrm{0})^2
     \label{eqn:3}
 \end{equation}
where $\theta_\mathrm{ijk}$ is the angle between two subsequent bond vectors, $\theta_\mathrm{0}=\pi$ is the reference angle and $k_{\theta}$ defines the bending stiffness. We set $k_{\theta}=k_{\mathrm{B}}T/\rm{rad}^2$ for flexible polymers and $k_{\theta}=20.0k_{\mathrm{B}}T/\rm{rad}^2$ for rod-like chains to reflect their increased rigidity. No bending potential is introduced for elastic shell bond vectors. 
The corresponding persistence length, $l_\mathrm{p}$, which characterizes the intrinsic stiffness of the chain, is given by $l_\mathrm{p}=k_{\mathrm{B}}Tl_\mathrm{b}/k_{\theta}$, where $l_\mathrm{b}$ is the average bond length. At temperature $T=1$, this model gives $l_\mathrm{b}\approx 0.97\sigma$~\cite{Kremer1990} and consequently  $l_\mathrm{p}=1\sigma$ for flexible polymers and $l_\mathrm{p}=19.4\sigma$ for semiflexible chains with $N_\mathrm{rod} = 8$ monomers, for instance, reflecting the much greater rigidity of the latter.


As a first step of preparation of simulation systems, we construct the elastic spherical shell as a single-layer meshwork composed of coarse-grained beads (see Fig.~\ref{fig1}a)~\cite{Attar2024}. All simulations use a shell consisting of $N_\mathrm{shell} = 22500$ beads, which are initially distributed randomly over a spherical surface with a radius of $R_0 = 51\sigma$. 
To prevent bead overlap and ensure a dense but uniform distribution of shell monomers, we introduce a distance parameter, $d$, which defines the minimum allowed separation between neighboring beads on the shell surface. This parameter is optimized to balance two constraints: avoiding steric clashes and allowing beads to fit into tight regions of the mesh (see Fig.~\ref{fig1}a).
Specifically, $d$ is chosen within the range
\begin{equation}
  0.60f < d < 1.0f
  \label{eqn:2}
\end{equation}
where $f$ is a minimization factor introduced to control the shell compaction during energy minimization. 
On average, each bead is connected to approximately 5.5 neighboring beads, typically forming between 5 and 8 bonds with its nearest neighbors. Bonding within the shell mesh is introduced by connecting neighboring beads that lie within a cutoff, $r$, such that
\begin{equation}
  0.66f < r < 0.9f
  \label{eqn:cutoff}
\end{equation}
If a minimum of $n_\mathrm{bond}=5$ bonds are not formed under these conditions (Eqn.~\ref{eqn:cutoff}), beads are allowed to bond with distant beads by expanding $r$ by 0.30.

In simulations, four long flexible polymers are confined within the elastic shell to generate osmotic pressure that counteracts excessive inward collapse of the shell and avoids any polymer-monomer linkages (see Fig.~\ref{fig1}b, pink chains). These polymers are organized into compact, grid-like blocks, each comprising $N_\mathrm{polymer} = 6002$ monomers, leading to a total polymer volume fraction of approximately $\phi \approx 10\%$. 
Each block is constructed by first placing an initial monomer at a random position within a radius of $34\sigma$ from the shell center (see Fig.~\ref{fig1}b). A random orientation is then assigned using directional angles constrained to ensure sufficient spacing between the polymer and shell, as enforced by the condition
  \begin{equation}
     \sqrt{(x_\mathrm{p}-x_\mathrm{s})^2 + (y_\mathrm{p}-
y_\mathrm{s})^2+(z_\mathrm{p}-z_\mathrm{s})^2}\geq 30\sigma
     \label{eqn:4}
 \end{equation}
where $(x_\mathrm{p}, y_\mathrm{p}, z_\mathrm{p})$ and $(x_\mathrm{s}, y_\mathrm{s}, z_\mathrm{s})$ denote the coordinates of the polymer and shell.
Polymer chains are then extended in both directions from the seed bead to form short linear chains, each 20 monomers long. This process is repeated iteratively in a plane to create a tightly packed 2D block of 10 chains (200 beads in total). Subsequent blocks are stacked atop the initial layer, with each layer spaced by $0.9\sigma$, slightly below the bond length $l_\mathrm{b}\approx \sigma$, to ensure close packing and controlled initial concentration.

In addition to flexible polymers, we also include semiflexible molecules confined within the elastic shell (see Fig.~\ref{fig1}a). Each rod-like molecule consists of $N_\mathrm{rod}$ monomers arranged linearly. To maintain semiflexible behavior, we keep the persistence-to-contour length ratio fixed at $l_\mathrm{p}/l = 2.5$ for all rod-like chains, where $l=(N_{\mathrm{rod}}-1)l_{\mathrm{b}}$. The number of semiflexible chains is varied across simulations to investigate their influence on the shape of the shell.
\begin{figure*}[ht]
\centering
  \includegraphics[width=0.85\linewidth]{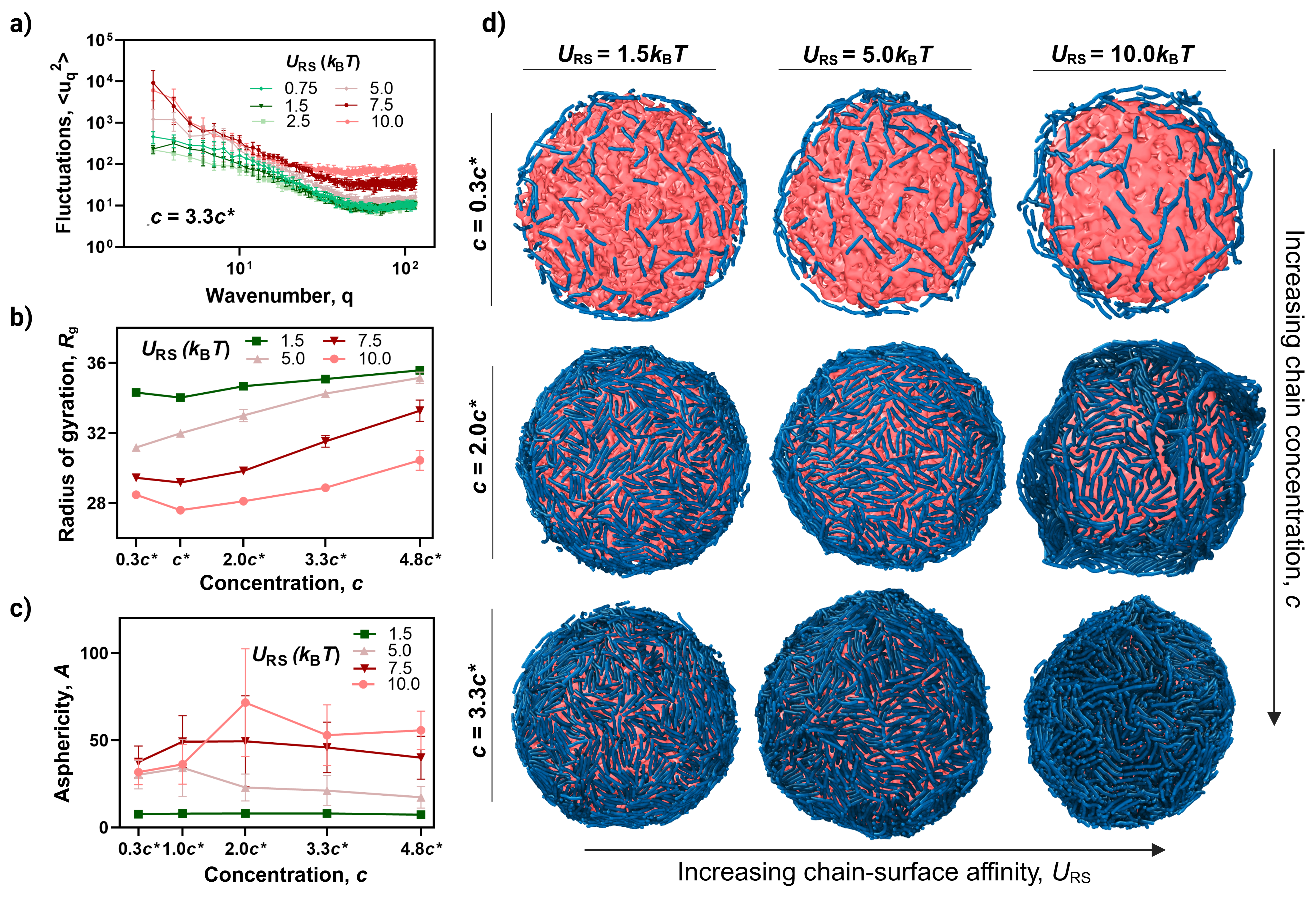}
  \caption{Effect of concentration of semiflexible chains, $c$, and chain-shell surface interaction strength, $U_\mathrm{RS}$, on the shape of the elastic shell. a) Shape fluctuation analysis for simulations for various chain-surface affinities, $U_\mathrm{RS}$. b) Radius of gyration, $R_\mathrm{g}$, and c) asphericity, $A$ as a function of concentration, $c$, at various chain-surface interaction strengths, $U_\mathrm{RS}$. d) Representative snapshots of the elastic shell shape with increasing $U_\mathrm{RS}$ (left to right) and increasing chain concentration, $c$ (top to bottom). The elastic shell is transparent for clear visualization. }
  \label{fig2}
\end{figure*}
%
%
%

Initially,  semiflexible polymer chains are randomly placed within the elastic shell, as illustrated in Fig.~\ref{fig1}b. To close large pores in the shell and prevent polymer or rod escape, the shell is isotropically shrunk by approximately $15\%$ through an energy minimization process. This minimization is carried out for $30\tau$ using an integration time step of $\Delta t = 0.001\tau$, with a damping coefficient of $0.1m\tau^{-1}$ (see Fig.~\ref{fig1}b, right panel).
Following minimization, the system is relaxed for $500\tau$ with only repulsive interactions active, allowing both the flexible polymers and semiflexible chains to equilibrate and lose memory of their initial configurations (see Fig.~\ref{fig1}b, center panel). After this, the main production run is performed for $1.25 \times 10^4\tau$  with a larger time step of $\Delta t = 0.005\tau$ (see Fig.~\ref{fig1}b, left panel). During this stage, attractive interactions between the rods and the shell are turned on to explore how surface adsorption of rods affects shell deformation. 

For a subset of simulations, we additionally replace FENE bonds of the shell defined in Eq.~\ref{eq:fene} with harmonic springs to generate a softer shell configuration. This final equilibration step is run for $10^6$ MD time steps. The harmonic bond potential is defined as
\begin{equation}
   U_{\text{Bond}}(r_\mathrm{ij}) = k (r_\mathrm{ij} - r_\mathrm{0})^2
   \label{eq:harmonic}
\end{equation}
where the equilibrium bond length is $r_\mathrm{0} = 1.2\sigma$ and the spring constant is $k = 5.0k_{\mathrm{B}}T/\sigma^2$, ensuring that shell connectivity is preserved without allowing chain escape.

All simulations are performed using the LAMMPS molecular dynamics package~\cite{Plimpton1995, Thompson2022}. Visualization and data analysis are conducted using OVITO~\cite{Stukowski2009} and Python, respectively.

We investigate how the collective behavior of surface-adsorbed semiflexible chains influences the shape of an elastic shell. Specifically, we explore the effects of varying three key parameters:
i) the concentration of semiflexible chains inside the shell, denoted by $c$;
ii) the ratio of the chain contour length to the shell radius, $l/R$; and
iii) the strength of chain–shell affinity, denoted by $U_\mathrm{RS}$.
The concentration, $c$, is controlled by adjusting the number of semiflexible polymers within the shell. 
We discuss concentration as a ratio of the bulk overlap concentration, $c^*$. The overlap concentration, $c^*$, is a polymer physical quantity, where individual rods begin to overlap and interact with one another~\cite{Rubinstein2003}. The overlap concentration for rod-like chains with $N_\mathrm{rod}$ = 8 monomers is $c^*=0.03\sigma^{-3}$ (see supplementary material Table S1).
We focus on chains of intermediate lengths, particularly in the regime where the contour length is comparable to the shell radius, i.e., $l \approx R$, with $R \approx 34\sigma$ throughout the study~\cite{Khadilkar2018, Naturephy2025} (see Fig.~\ref{fig1}b). 
Note that the initial size of the shell, prior to minimization, is $R_\mathrm{0} = 51\sigma \approx 1.5R$ (see Fig.~\ref{fig1}b). The chain length, $l$, is varied from $4\sigma$ to $64\sigma$ by varying the number of beads per rod-like chain, $N_\mathrm{rod}$, corresponding to a range of $0.1 \leq l/R \leq 2$.
The chain–shell interaction strength, $U_\mathrm{RS}$, is modulated by changing the interaction parameter $\epsilon$ in Eqn.~\ref{eq:lj}. Note that all beads within a semiflexible polymer are identical and interact uniformly with the shell interior, ensuring that the entire chain participates equally in surface interactions.

%
\subsection{\label{m3}Shape analysis of elastic shell}

The equilibrated shapes of the elastic shell are described by the radius of gyration ($R_\mathrm{g}$), asphericity ($A$), and shape anisotropy ($\kappa$) (see supplementary material Fig. S1), all derived from the gyration tensor ($\mathbf{S}$)~\cite{Janke2013}, as well as by shape fluctuations (see Fig.~\ref{fig1}c)~\cite{Attar2024}.  
%
The gyration tensor $\mathbf{S}$ of an elastic shell is defined as 
\begin{equation}
 S_{\alpha\beta} = \frac{1}{N_{\mathrm{shell}}} \sum_{i=1}^{N_{\mathrm{shell}}} (r_{i,\alpha} - r^{\mathrm{cm}}_{\alpha})(r_{i,\beta} - r^{\mathrm{cm}}_\beta)   
\end{equation}
where $r_{i,\alpha}$ is the $\alpha$-th component of the position of bead $i$, and $r^{\mathrm{cm}}_\alpha$ is the corresponding component of the center of mass with $\alpha,\beta=x,y,z$.  
Transforming the gyration tensor, $\mathbf{S}$, to its principal axis system yields a diagonal form
\begin{equation}    
\mathbf{S} = \text{diag}(\lambda_1, \lambda_2, \lambda_3) 
\end{equation}
where the eigenvalues are sorted in descending order: $\lambda_1 \geq \lambda_2 \geq \lambda_3$.
The squared radius of gyration is given by the trace of $\mathbf{S}$, which corresponds to the sum of its eigenvalues
\begin{equation}
    R_\mathrm{g}^2 = \text{Tr}(\mathbf{S}) = \lambda_1 + \lambda_2 + \lambda_3
\end{equation}
%
%
%

The asphericity, $A$, quantifies the deviation of the shell shape from a perfect sphere and is defined as~\cite{Janke2013}
\begin{equation}
    A = \lambda_1 - \frac{1}{2} (\lambda_2 + \lambda_3) 
\end{equation}
An asphericity value of $A \approx 0$ corresponds to a nearly spherical shape, while higher values (e.g., $A \geq 50$) indicate increasingly distorted or elongated configurations (see Fig.~\ref{fig1}c). Asphericity has been widely used to characterize shape fluctuations of polymers under confinement or within vesicles, etc.~\cite{Naturephy2025, May2013, Liu2008, Liu2024}. 


To quantify shell shape fluctuations, $\langle u_q^2 \rangle$, we compute Fourier modes from a 1$\sigma$-thick shell slice taken through the center along the $x$-axis at each time step~\cite{Attar2024, Banigan2017, Patteson2019}. For each bead in the slice, we calculate its angular deviation from the average radial distance. A Fast Fourier Transform (FFT)~\cite{Harris2020} is then applied to the lowest $q = 230$ modes, and the results are time-averaged over the second half of the simulation. This approach captures shape fluctuations independently of the overall shell size (see Fig.~\ref{fig1}c).

In the results discussed below, the radius of the elastic shell, $R \approx 34\sigma$, persistence length-to-contour length ratio, $l_\mathrm{p}/l = 2.5$, and volume fraction of flexible polymers, $\phi \approx 10\%$, are fixed, unless otherwise stated (see Fig.~\ref{fig1}a, b). The elastic shell is often transparent in the simulation snapshots shown below for a clear visualization of the semiflexible chain behavior at the shell surface. In most MD simulations, more than $75\%$ of the polymer chains are localized to the interior surface of the elastic shell.

%
%
\begin{figure*}[ht]
\centering
  \includegraphics[width=\linewidth]{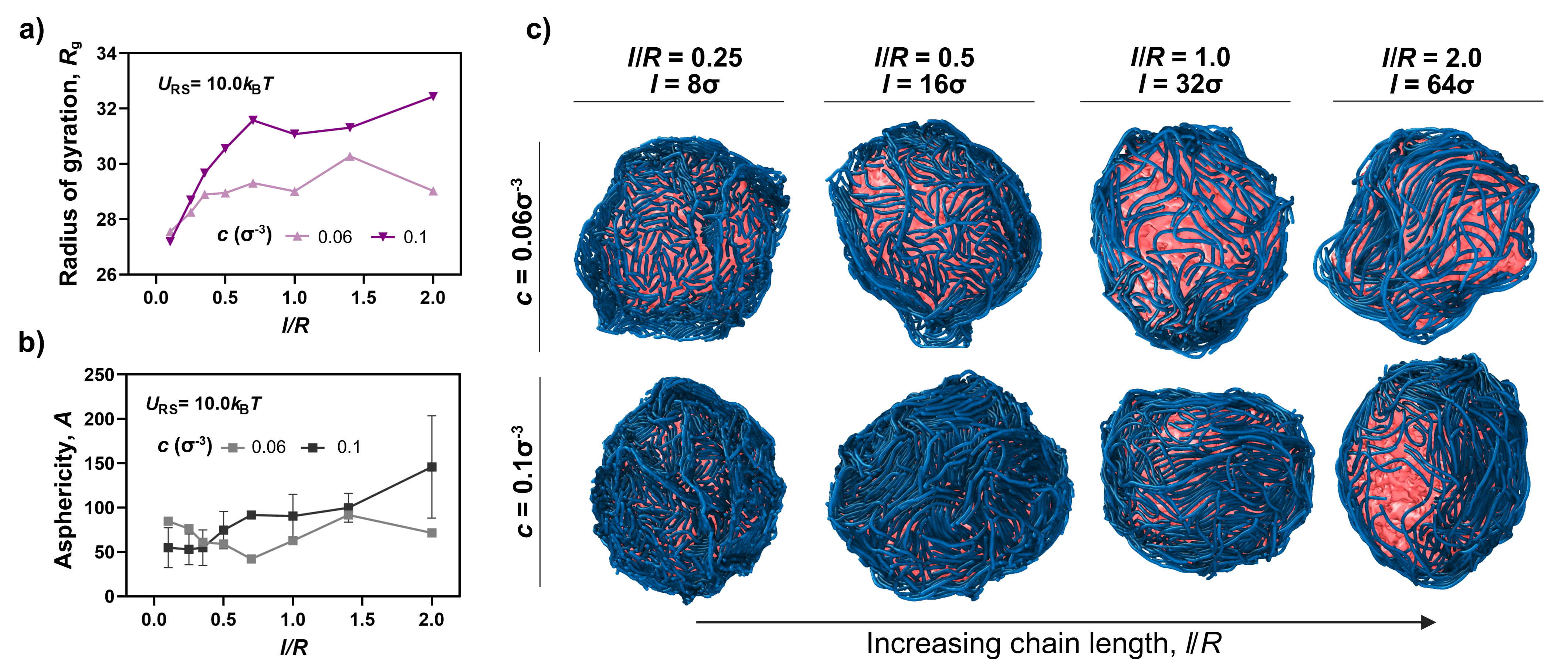}
  \caption{Effect of chain contour length, $l$, on the shape of the elastic shell under strong chain-surface interactions, $U_\mathrm{RS} = 10.0k_\mathrm{B}T$. a) Radius of gyration, $R_\mathrm{g}$, and b) asphericity, $A$ as a function of varying $l/R$ for $c > c^*$. c) Simulation snapshots for $l/R$ = 0.25, 0.5, 1.0, and 2.0 (i.e., $8\sigma \leq l \leq 64\sigma$) at $c$ = $0.06\sigma^{-3}$ and $0.1\sigma^{-3}$.}
  \label{fig3}
\end{figure*}
\section{\label{results}Results and discussion}
Our study attempts to understand how \textit{surface-adsorbed}, semiflexible polymer chains under varying attractive interactions with the shell surface can affect the morphology and shape fluctuations of a pressurized elastic shell. Therefore, we vary chain-surface attraction, $U_\mathrm{RS}$, chain concentrations, $c$, chain length-to-shell radius ratios, $l/R$, and shell structure.  Our MD trajectories are analyzed to characterize the shape changes by using metrics such as fluctuation spectrum, asphericity, anisotropy, and average dimension of the shell (see Fig.~\ref{fig1}c, and supplementary material Fig. S1). For clarity, the terms ``chain'' and ``polymer'' will be used to refer to the semiflexible polymers throughout this work.
\subsection{\label{r1} Strong localization of chains to the interior surface leads to shape distortions}
We first analyze the effect of the peripheral localization of semiflexible chains on the shape of the elastic shell in our simulations. To do so, we vary the interaction strength between the chains and the inner surface of the elastic shell, $U_\mathrm{RS}$. 
We choose a relatively low contour length-to-shell radius ratio (i.e., $l/R = 8/34 \approx 0.25$) to avoid the buckling or strong bending of the adsorbed chains due to the surface curvature (see Fig.~\ref{fig2}, and supplementary material Fig. S2). At low to intermediate polymer-shell interaction strengths (i.e., comparable to or higher than thermal energy,  $U_\mathrm{RS} = 1.5$ and $5.0 k_\mathrm{B}T$), the shape of the elastic shell appears to be spherical regardless of the chain concentration (see Fig.~\ref{fig2}d, first two columns). On the contrary, as the attraction strength between the surface and polymers is increased (i.e., $U_\mathrm{RS} >  5.0k_\mathrm{B}T$), the adsorbed chains become largely immobile on the surface, and the elastic shell locally morphs around the chains, leading to stable, non-spherical geometries (see Fig.~\ref{fig2}d, last column).

Analyses of the shape spectrum (see Methods~\ref{m3}, Fig.~\ref{fig1}c) also confirm that excess binding of chains to the surface leads to systematically higher fluctuation amplitudes in all wavelengths, suggesting that shape distortions are dominant from the scales of several shell beads up to the scale of the entire shell (see Fig~\ref{fig2}a).
%

\subsection{\label{r2}Shape distortions exhibit a weak concentration dependence}
Having demonstrated the effect of chain-surface interactions, $U_\mathrm{RS}$, on the shape of the elastic shell, we next explore whether these trends depend on the concentration of semiflexible chains, $c$, confined inside the elastic spherical confinement. To explore this, we choose concentrations above, below, and equal to the overlap concentration (i.e., $0.3c^* \leq c \leq 4.8c^*$) (see Introduction \ref{intro}, and supplementary information). Then, we systematically vary $U_\mathrm{RS}$ for each concentration (see Fig~\ref{fig2}d). 

At low concentrations, $c \leq c^*$, the shape of the elastic shell appears largely spherical, regardless of polymer-shell attraction strength (see Fig.~\ref{fig2}d, first row).  On the contrary, at high concentrations (i.e., $c > c^*$), the spherical shell exhibits a more distorted morphology with strong chain localization, $U_\mathrm{RS} = 10.0k_\mathrm{B}T$(see Fig.~\ref{fig2}c,  red data). We also observe significant wrinkles on the shell surface, manifested by increasing fluctuation intensity in the shape spectrum (see Fig~\ref{fig2}d, second and third row, and supplementary material Fig. S2). 

The asphericity, $A$, is more affected if the chains are bound to the surface strongly. The shape fluctuations also lead to larger error bars in asphericity parameters at high concentrations, pushing the numerical value of asphericity to $A \approx 100$ (see Fig.~\ref{fig2}c, $2.0c^*\leq c \leq 4.8c^*$).

Another immediate effect of concentration under strong-localization conditions observed in simulation is that as concentration increases, the radius of gyration of the shell, $R_\mathrm{g}$, increases. This occurs for all $U_\mathrm{RS}$ but is more pronounced when chain localization is strong ($U_\mathrm{RS} >  5.0k_\mathrm{B}T$) (see Fig.~\ref{fig2}b, and supplementary material Fig. S3a). This is due to crowding of the chains on the surface, further expanding the shell despite there being less translational entropy (and related osmotic pressure) in the shell.

Notably, the shape anisotropy parameter, $\kappa$, also increases by an order of magnitude as $U_\mathrm{RS}$ increases (see Methods \ref{m3}, and supplementary material Fig. S3b). This could be due to wrinkles in the shell that emerge at high polymer-shell interactions (i.e., at $U_\mathrm{RS} = 7.5$ and $10.0k_\mathrm{B}T$) (see Fig.~\ref{fig2}d, last column, and supplementary material Fig. S2).

Overall, our calculations show that if the chain concentration is low enough (i.e., $c \leq c^*$), the adsorption of semiflexible polymers has a weak effect on the shell shape, leading to more spherical geometries (see Fig.~\ref{fig2}d, first row). As the concentration increases, $c \geq c^*$, the chain-surface interactions, $U_\mathrm{RS}$, play a dominating role. As a result, the shell exhibits stable, distorted shapes under strong localization (see Fig~\ref{fig2}d, last column, and supplementary material Fig. S4 and S5).

\subsection{\label{r3} Strong localization of chains on the shell causes shape fluctuations irrespective of chain length}

Next, we examine whether the shape distortions discussed above depend on the chain length, $l$. In previous sections (\ref{r1} and~\ref{r2}), the chain length to shell radius ratio is fixed at $l/R = 8/34 \approx 0.25$. Here, we vary $l/R$ by altering chain length, $l$, within the range $4\sigma \leq l/R \leq 64\sigma$, while keeping the elastic shell radius unchanged at $R\approx 34\sigma$ (see Fig.~\ref{fig1}b). The chain-surface interaction strength is adjusted to $U_\mathrm{RS} = 10.0k_\mathrm{B}T$ to obtain strong surface adsorption. 
The chain concentration is kept high such that $c > c^*$ to observe shape distortions due to strong localization (see Fig.~\ref{fig2}). Given that the overlap concentration, $c^*$, varies with the chain length, we keep absolute concentration at $c=0.06\sigma^{-3}$ and 0.1$\sigma^{-3}$, corresponding to $c > c^*$ for all chain lengths (see supplementary material Table S2).  We choose $l/R$ values at which the chain contour length is comparable to or smaller than the radius of the elastic shell, $R$ (i.e., $l/R \leq 2$). 
Note that as we increase the chain length, we preserve the ratio $l_\mathrm{p}/ l = 2.5$ for all cases (see Methods \ref{m1} and Fig.~\ref{fig1}a). Thus, the persistence length, $l_\mathrm{p}$, also increases and exceeds the radius of the shell, $R$.

For chain concentrations considered here (i.e., $c > c^*$ for all chain lengths),  the elastic shell exhibits strong shape distortions (see Fig.~\ref{fig2}c). That is, fluctuation amplitudes and shape anisotropy parameters are independent of $l/R$ (see supplementary material Fig. S6 and S7). Nevertheless, visual inspection shows that as $l/R$ increases, the chains distribute non-uniformly on the surface of the elastic shell (see Fig.~\ref{fig3}c, and supplementary material Fig. S8). For instance, while a chain concentration of $c \approx 0.1\sigma^{-3}$ is high enough to densely cover the surface, empty, chain-free sections appear on the surface for longer chain lengths (i.e., $l/R$ = 1.0 and 2.0 in Fig.~\ref{fig3}c, second row, and see supplementary material, Fig. S8 for $l/R$ = 1.4). 
%

To further characterize these abnormal shape morphologies, we plot the time-averaged radius of gyration and asphericity of the elastic shell as a function of $l/R$ for a range of ratios, $0.1 \leq l/R \leq 2$ (see Fig.~\ref{fig3}a-b). The elastic shell radius of gyration increases with increasing $l/R$  (see Fig.~\ref{fig3}a). 
Given that longer chains (i.e., $l/R \geq 1.0$)  buckle inside the shell, there are swollen, chain-free regions on the surface, which could be related to this size increase. The radius of gyration does remain below $ R\approx34\sigma$ (i.e., the shell does not swell beyond its size at purely repulsive chains-surface interactions) (see Fig.~\ref{fig1}b).
Given the strong localization of the chains, all cases in Fig.~\ref{fig3}b also exhibit high asphericity as well (i.e., 50 $\leq A$ < 125) (see Fig.~\ref{fig3}b). Surprisingly, the asphericity is higher for higher absolute chain concentration (i.e., $c \approx 0.1\sigma^{-3}$) at $l/R>0.5$. We speculate that this could be due to non-uniform chain distribution that leads to parallel chain orientation as $l/R$ increases~\cite{Girard2024}(see Fig.~\ref{fig3}c, second row).

Hence, we conclude that strong localization can distort the shape of an elastic shell irrespective of chain length (see Fig.~\ref{fig3}c). These distortions might not be visible in the fluctuation spectra but can alter asphericity and the size of the shell (see Fig.~\ref{fig2} and ~\ref{fig3}).

%
%
%
\subsection{\label{r4} Nematic domains of chains emerge with increasing chain length under weak localization}

In the previous sections, we investigate the effect of strong chain localization on the shape of our elastic shell. Under strong localization, the chains distort the shell shape significantly (see Fig.~\ref{fig2} and~\ref{fig3}). Nevertheless, strong localization does not allow the formation of distinct phases on the shell surface. Thus, next, we look into the shape alterations that could emerge under weak surface adsorption of semiflexible polymers as a result of nematic phase formation of many chains. 

%
\begin{figure}[h]
\centering
  \includegraphics[width=\linewidth]{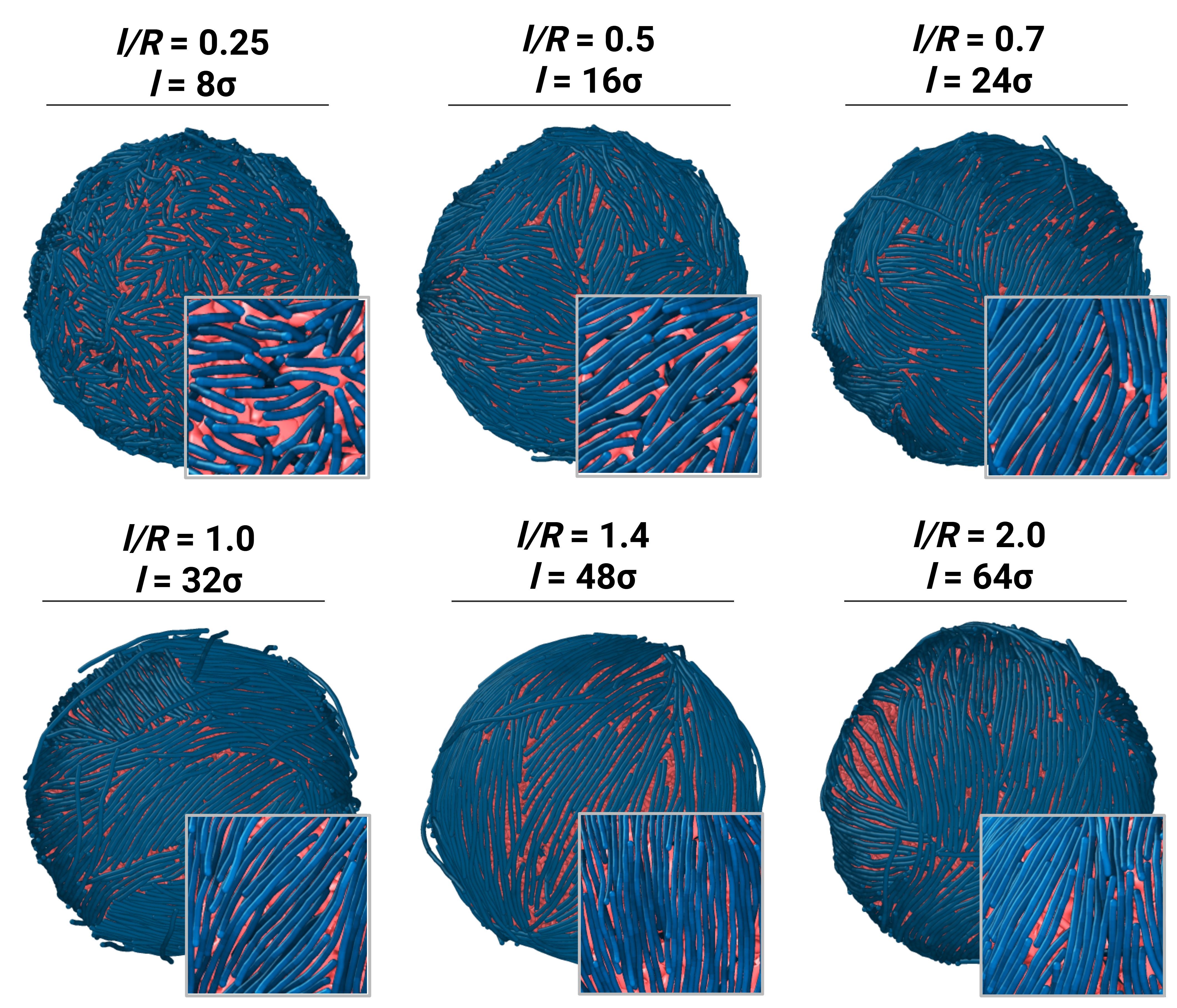}
  \caption{Effect of chain length, $l/R$, on organized domain formation of rod-like polymers confined within an elastic shell. Rod-shell interaction strength is weak, $U_\mathrm{RS} = 1.5k_\mathrm{B}T$. Chain concentration is fixed to high, $0.1\sigma^{-3}$, $c>c^*$. Chain length, $l$, is varied such that $8\sigma \leq l \leq 64\sigma$.}
  \label{fig4}
\end{figure}
%


Fig.~\ref{fig2} shows that, when the chains are relatively short (i.e., $l/R=0.25$), a weak chain-surface interaction strength (i.e., $U_\mathrm{RS} \approx k_\mathrm{B}T$)  is enough to localize most of the chains to the shell surface (see Fig.~\ref{fig2}d, first column). Nonetheless, this localization does not lead to detectable deviations in the shell shape as indicated by: i) consistently low fluctuation amplitudes (see Fig.~\ref{fig2}a); shell radius of gyration close to the initial shell size (i.e., $R_\mathrm{g} \approx R=34\sigma$)(see Fig.~\ref{fig2}b); and low asphericity, $A < 25$ (see Fig.~\ref{fig2}c); for all chain concentrations considered in this work. 

Similar to our analyses in Section~\ref{r3}, we characterize the effect of increasing $l/R$ for high absolute chain concentration, $c \approx 0.1\sigma^{-3}$, where the surface is completely covered at $c > c^*$ (see Fig.~\ref{fig4}, and supplementary material Fig. S9). The chain-surface interaction is kept constant at a relatively low value of $U_\mathrm{RS} = 1.5k_\mathrm{B}T$. For short chains (i.e., $l/R \approx 0.25$, $0.5$), chains coat the surface with no visually uniform orientation (see Fig.~\ref{fig4}, $l = 8\sigma$ and $16\sigma$). As we increase $l/R$ to $0.7$ and $1.0$, locally ordered nematic-phase domains emerge on the surface (see Fig.~\ref{fig4}, $l = 24\sigma$ and $32\sigma$). This is consistent with previous studies on semiflexible chains in rigid spherical confinements ~\cite{Nikoubashman2017, Nikoubashman2021, Binder2018, Milchev2017, Milchev2018}. Increasing $l/R$ further to higher values (i.e., $l/R \approx$ $1.4$ and $2.0$) results in a more unidirectional arrangement of the polymer chains (see Fig.~\ref{fig4}, $l = 48\sigma$ and $64\sigma$). 

To summarize, under weak localization, where the shape of the elastic shell is largely spherical, increasing chain length leads to the formation of nematic phases at high concentration (i.e., $c>c^*$~\cite{Hameed2024, Milchev2017}. These domains become more unidirectional as the rod length exceeds the shell radius (see Fig.~\ref{fig4}, second row).

\begin{figure*}[ht]
\centering
  \includegraphics[width=\linewidth]{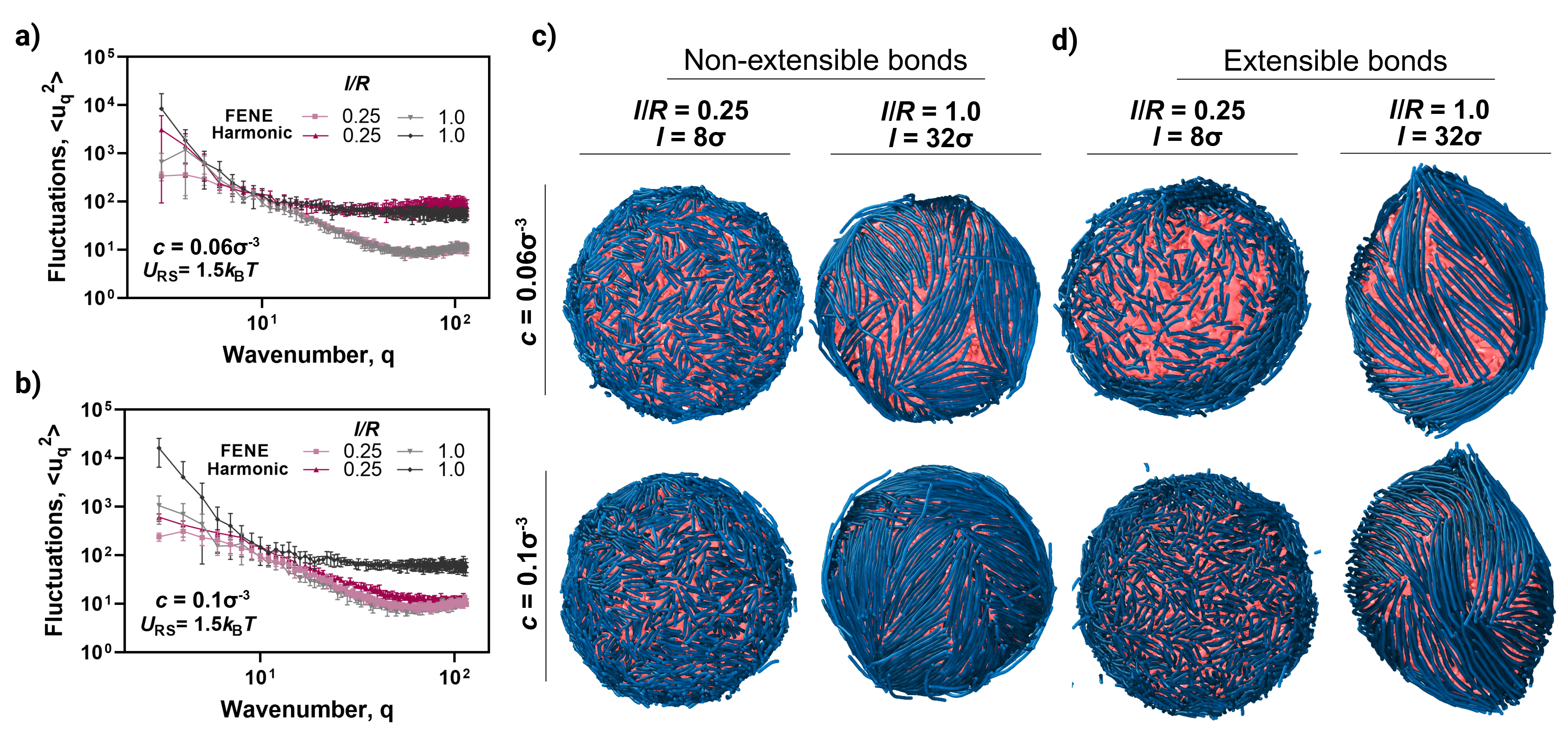}
  \caption{Formation of elliptic shell shapes with increasing chain length, $l/R$, and switching to extensible harmonic bonds under weak rod-shell interaction strength, $U_\mathrm{RS} = 1.5k_\mathrm{B}T$. a-b) Shape fluctuation comparison for non-extensible FENE vs. extensible harmonic shell bonds for a) $c \approx 0.06\sigma^{-3}$, and b) $c \approx 0.1\sigma^{-3}$. Simulation snapshots comparing c) FENE vs. d) harmonic elastic shell configurations for $l/R$ = 0.25 and 1.0.  }
  \label{fig5}
\end{figure*}
\subsection{\label{r5} Weak chain localization to the shell leads to elongated shapes for softer shells}
So far, to connect the shell beads, we have used a model bond potential referred to as Finitely Extensible Non-linear Elastic (FENE) bonds (Eqn.~\ref{eq:lj}). We next test whether making the shell more deformable by replacing these non-extensible bonds with an extensible harmonic potential (Eqn.~\ref{eq:harmonic}) would affect the shell shapes (see Methods~\ref{m1}). 
Consistent with previous studies~\cite{Milchev2017, Milchev2018}, we chose chain lengths (i.e., $l/R$ values) representing both disordered and ordered (i.e., more nematic) chain orientations on the surface (i.e., $l/R \approx 0.25$ and $1.0$ in Fig.~\ref{fig4}) for high chains concentrations (i.e., $c >c^*$ with an absolute concentrations $c \approx 0.06$ and $0.1 \sigma^{-3}$) (see Fig.~\ref{fig5}). 

Making the bonds extensible leads to highly distorted shapes, irrespective of chain length (see Fig.~\ref{fig5}d). 
%
We visually observe an irregular spherical shape for $l/R \approx 0.25$, and an elliptic shape for $l/R \approx 1.0$ elongated in the directions. These shapes are reminiscent of lipid vesicles where rod-like microtubules were strongly confined~\cite{Bausch2014}  (see Fig.~\ref{fig5}d, first row). The distorted morphologies we observe also manifest themselves as higher fluctuation amplitudes than FENE-bonded spherical shells, when $c = 0.06\sigma^{-3}$ (see Fig~\ref{fig5}a). 

Surprisingly, when we slightly increase chain concentration to $c = 0.1\sigma^{-3}$ from $c = 0.06\sigma^{-3} $, where surface-adsorbed chains cover almost the entire surface, shape distortions become $l/R$ dependent (see Fig.~\ref{fig5}b). For shorter chains (i.e., $l/R \approx 0.25$), the fluctuations for both FENE and harmonic bonded shells are comparable since the chains organize randomly on the elastic-shell surface (see Fig.~\ref{fig4}c, second row).
On the contrary, with longer chains (i.e., $l/R \approx 1.0$), where locally nematically ordered chain domains appear, the harmonic shell undergoes a transition to an elliptic shape  (see Fig.~\ref{fig4} and ~\ref{fig5}c, second row).
These shapes exhibit a higher fluctuation amplitude than their FENE-bonded counterparts. 

Overall, our simulations reveal that weak chain localization to the elastic shell, $U_\mathrm{RS} = 1.5k_\mathrm{B}T$, leads to a orientational ordering with increasing chain length, which, in turn, causes strong morphological distortions of the shell in a chain length-dependent manner, provided that bonds of the shell are deformable enough to extend (see Fig.~\ref{fig4},~\ref{fig5}, and supplementary material Fig. S10). Such phase behavior and shell elongation are not observed when chains are strongly localized to the shell (i.e., $U_\mathrm{RS} = 10.0k_\mathrm{B}T$). Instead, those shells collapse when the shell has extensible harmonic bonds (see supplementary material, Fig. S4). This could be due to an imbalance of osmotic pressure between chain-free and chain-dense regions (see Fig.~\ref{fig3}).

\section{Conclusions}

In this work, using a model system and MD simulations, we study the effects of the surface adsorption of many (i.e., above overlap concentration) semiflexible polymers on the morphology of an elastic shell. 
We focus on how the shape of a pressurized shell is deviated from a perfect sphere as a result of strong or weak localization of many polymer chains, depending on the chain lengths and concentrations. This 3-component system (see \nameref{methods}) could be considered a crude vesicle or nucleus model~\cite{Attar2024, Bausch2016, Okano2018, Naturephy2025} (Fig.~\ref{fig1}). Our calculations showed that formation of distinct surface phases (nematic or isotropic phases) and/or attraction affinity of chains towards the shell surface can lead to a variety of shapes and shape fluctuations with various amplitudes, including egg-like shapes and bleb-like protrusions.
 

Our results elucidate how surface adsorption of semiflexible polymers may alter the shape of elastic shells,  such as lipid vesicles or hydrogel shells, via several mechanisms; i) shell wrinkling and stable morphology due to excess chain localization (see Fig.~\ref{fig2} and ~\ref{fig3}); and ii) organized domain formation under weak localization (see Fig.~\ref{fig4}), correlated to shell elongation when the shell deformability increases (see Fig.~\ref{fig5}). 
 
We demonstrate that an elastic shell can undergo significant shape fluctuations under strong localization of the chains on the interior surface of the shell (i.e., strong binding affinity to the surface) (Fig.~\ref{fig2}). As the polymer-surface interaction strength increases, chains may become immobile at the shell surface, further stabilizing the shell, which supports the notion that scaffolds composed of such polymer can provide mechanical stability~\cite{Baumann2020, Liu2008, Fakhri2014} (see Fig.~\ref{fig3}). These shape fluctuations occur above the bulk overlap concentration, $c > c^*$ (see Fig.~\ref{fig2}), and are largely independent of chain length (see Fig.~\ref{fig3}). Under weak localization conditions (i.e., binding affinity is on the order of thermal energy),  semiflexible chains exhibit nematic ordering on the surface with increasing chain length, $l/R$, when chain concentration within the confinement is high, $c > c^*$ (see Fig.~\ref{fig4}). This validates previous studies with rigid confinement, restating that inter-chain interactions are not required to observe local nematic ordering for semiflexible polymers~\cite{Binder2018, Milchev2018, Milchev2017}. This ordering does not distort the shape of the elastic shell if the localization strength is weak. 
Nevertheless, as the mechanical softness of the shell is increased, the shell is more prone to undergo shape anomalies (i.e.,  elongated, egg-like shell shapes), suggesting the competition between shell mechanics and chain elasticity (see Fig.~\ref{fig5}). 
Notably, these oval shell shapes become more pronounced with increasing chain length and concentration (see Fig.~\ref{fig5}d), consistent with previous studies under strong confinement limit~\cite{Bausch2014}.

Overall, our study provides a qualitative perspective on the behavior of surface-adsorbed semiflexible polymers by including shell elasticity as an additional competitive metric. Possible extensions to this work could include inter-chain interactions and/or meshless membrane models to model the shell in a more biologically accurate manner~\cite{Fu2017, Attar2024, PaulJ2013, Attar2025}.


\section{\label{weaknesses} Weaknesses of the study}

The elastic shell in our simulations is composed of a meshwork of beads, unlike membranes and vesicles, which are structurally heterogeneous and fluid-like. While we explore softer shell configurations via harmonic bonds between the beads of the elastic shell (see Fig.~\ref{fig5}), the elasticity remains uniform, unlike biological membranes, where it may depend on lipid composition, protein interactions, and external forces. etc. 

We also note that the model solely focuses on the surface adsorption of semiflexible polymers at the interior of the shell, with only repulsive interactions between them. On the contrary, self-assembled biological networks often consist of sticky ends that could be modeled via intermolecular interactions~\cite{Hameed2024, Yang2023}.
%
The scope of this study is also limited to semidilute concentration of semiflexible polymers with intermediate chain lengths (i.e.,  $0.1 \leq l/R \leq 2$), given the unexplored nature of length scales, where $l \approx R$. The ratio of persistence length-to-contour length is also fixed to $l_\mathrm{p}/l = 2.5$, meaning that we do not test the effect of varying bending rigidity on the shape of the elastic shell, though the flexibility can affect the shell shape only under strong shell affinity and polymer self-affinity combined~\cite{Attar2024}. Nevertheless, given that biopolymers of interest in most biological systems, i.e., actin filaments in the cytoskeleton vs. lamins lining the nucleus, may exhibit a distribution of bending rigidities. Last but not least, the activity of polymers (their time- or confinement-dependent length), which can lead to more exotic morphologies~\cite{Bausch2014,Bausch2016}, could be the subject of future studies.


\section*{Conflicts of interest}
There are no conflicts to declare.

\section*{Data availability}
All codes, structure files, and input parameters are available online at https://github.com/erbaslab


\section*{Acknowledgments\protect\\} 
This research was supported by TUBITAK, the Scientific and Technological Research Council of Turkey [Grant No. 124N935].




\bibliography{main} 
\bibliographystyle{main} 

\end{document}


\begin{titlepage}
    \thispagestyle{empty} 
    \centering
    \vspace*{\fill}
    \textbf{\Large Supplementary data and figures for:   \\   Shape spectra of  elastic shells with surface-adsorbed semiflexible polymers} \\
    \vspace{0.5cm} 
    Hadiya Abdul Hameed,\textsuperscript{1} and Jaroslaw Paturej,\textsuperscript{2} and Aykut Erbas\textsuperscript{1,2} \\
    \vspace{0.5cm}
    \begin{minipage}{0.7\textwidth} 
        \centering
        \textit{\textsuperscript{1}UNAM - National Nanotechnology Research Center and Institute of Materials Science and Nanotechnology, 
        Bilkent University, Ankara 06800, Turkey \\}
        \textit{\textsuperscript{2}Institute of Physics, University of Silesia at Katowice, Chorzów, 41-500, Poland}
    \end{minipage}
    \vspace*{\fill}
    \newpage 
\end{titlepage}

\maketitle
\newpage

\section{Semiflexible chain concentration, $c$}
The concentration, $c$, of coarse-grained semiflexible polymers inside the elastic shell is varied by changing the number of chains, $N$, within the elastic shell. 

The absolute concentration, $c$ in $\sigma^{-3}$, is calculated as
\begin{equation}
   c =  \frac{NN_\mathrm{rod}}{V}
\end{equation}
where $N_\mathrm{rod}$ is the number of monomers per chain.

We also denote chain concentration in terms of bulk overlap concentration, $c^*$
\begin{equation}
        c^* = \frac{N_\mathrm{rod}}{v} 
  \label{eq:c*}
\end{equation}
where $v \approx l\sigma^3$ is the pervaded volume of each chain. 

For $N_\mathrm{rod} = 8$, the overlap concentration is as follows 

\begin{equation}
   \begin{aligned}
        c^* &= \frac{8}{\frac{4}{3} \pi (4 \sigma)^3} = 0.02984155183 \approx 0.03 \sigma^{-3}
   \end{aligned}
  \label{eq:c1*}
\end{equation}

\begin{equation*}
   \begin{aligned}
   \text{where} \quad &
       N_\mathrm{rod} = 8, \quad &
       v = \frac{4}{3} \pi r^3, \quad & r = \frac{8\sigma}{2} = 4\sigma
   \end{aligned}
\end{equation*}

The concentrations probed for short chains (i.e., $l/R = 8/34 \approx 0.25$) are shown in Table~\ref{tab:tab1}. These correspond to the concentration range, $0.3c^*\leq c \leq 4.8c^*$.

\begin{table}[htbp]
  \centering
  \caption{Number of rod polymers, \textit{N}, and concentrations, \textit{c}, for $N_\mathrm{rod}$ = 8 .}
  \label{tab:tab1}
    \begin{tabular}{ccc}
      \textit{$N$} & $c(\sigma^{-3})$ & $c/c^*$ \\
      \hline
      250   & 0.01 & 0.3 \\
      750   & 0.03   & 1.0\\
      1250  & 0.06  & 2.0 \\
      2000  & 0.097   & 3.3 \\
      3000  & 0.15   & 4.8\\
    \end{tabular}
\end{table}
\newpage
Since the pervaded volume of the chain, $v$, increases with decreasing chain length, $l = N_\mathrm{rod}\sigma$, the overlap concentration, $c^*$, decreases. For example, for $N_\mathrm{rod} = 32$, the overlap concentration is

\begin{equation}
   \begin{aligned}
        c^* &= \frac{32}{\frac{4}{3} \pi (16 \sigma)^3} \approx 0.002 \sigma^{-3}
   \end{aligned}
  \label{eq:c2*}
\end{equation}

\begin{equation*}
   \begin{aligned}
   \text{where} \quad &
       N_\mathrm{rod} = 32, \quad &
       v = \frac{4}{3} \pi r^3, \quad & r = \frac{32\sigma}{2} = 16\sigma
   \end{aligned}
\end{equation*}

The overlap concentrations, and respective absolute concentrations in terms of $c^*$ for various $l/R$ discussed in the main text are shown in Table~\ref{tab:tab2}.

\begin{table}[htbp]
  \centering
  \caption{Overlap concentrations, and normalized concentrations for various chain lengths, where $R\approx 34\sigma$. }
  \label{tab:tab2}
    \begin{tabular}{ccccc}
      \textit{$N_\mathrm{rod}$} & $l/R$ & $c^*$ & $c/c^* (c = 0.06\sigma^{-3})$ & $c/c^* (c = 0.1\sigma^{-3})$\\
      \hline
     8   & 0.25 & 0.03 & 2.0 & 3.3 \\
      16 & 0.5  & 0.007   &  8.0 & 13.4\\
      24 & 0.7 & 0.003 &  18.1 & 30.1\\
     32 & 1.0 & 0.002   & 32.2 & 53.7 \\
      48 & 1.4 & 0.0008   &  72.4 & 120.6\\
      64 & 2.0 & 0.0005   &  128.7 & 214.5\\
    \end{tabular}
\end{table}

\section{Shape anisotropy, $\kappa$}
The relative shape anisotropy, $\kappa$, is a dimensionless measure that captures both the symmetry and dimensionality of a shape~\cite{Janke2013}. It is defined as
\begin{equation}
\kappa = 1 - \frac{3(\lambda_1\lambda_2 + \lambda_2\lambda_3 + \lambda_3\lambda_1)}{(\lambda_1 + \lambda_2 + \lambda_3)^2}.
\end{equation}
where $\lambda_1$, $\lambda_2$, and  $\lambda_3$ are eigenvalues of the gyration tensor of the elastic shell, $S$ (see main text).

The parameter  $\kappa$ ranges from 0 to 1: it equals 0 for perfectly symmetric (e.g., spherical) conformations and approaches 1 for elongated, linear structures. 
For planar symmetric configurations, $\kappa$ typically converges to 1/4.

\begin{figure}[h!]
  \centering
  \includegraphics[width=0.6\textwidth]{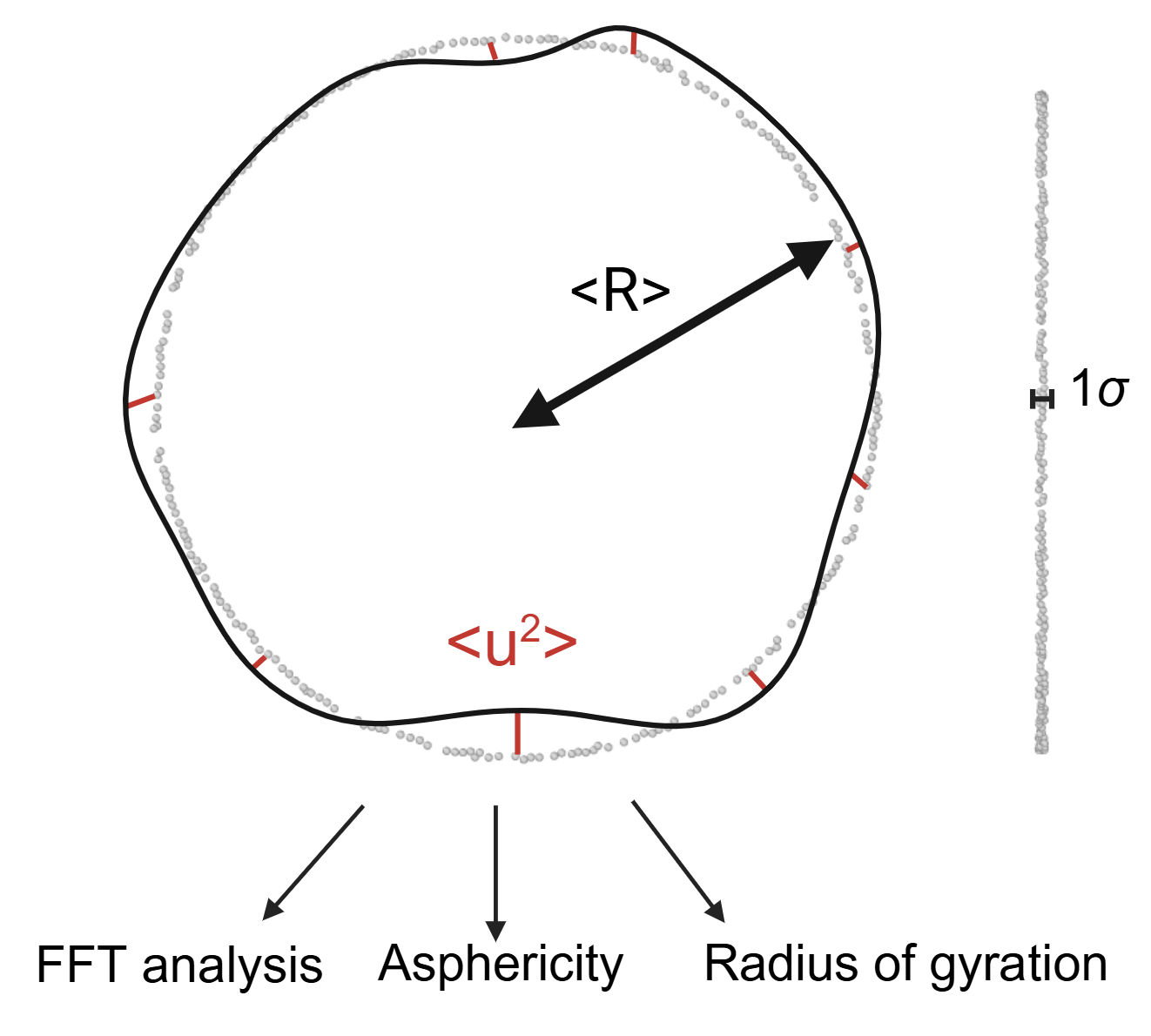}
  \caption{Analysis methods used to characterize shell shape anomalies in the main text. }
\end{figure}

\begin{figure}[h!]
  \centering
  \includegraphics[width=\textwidth]{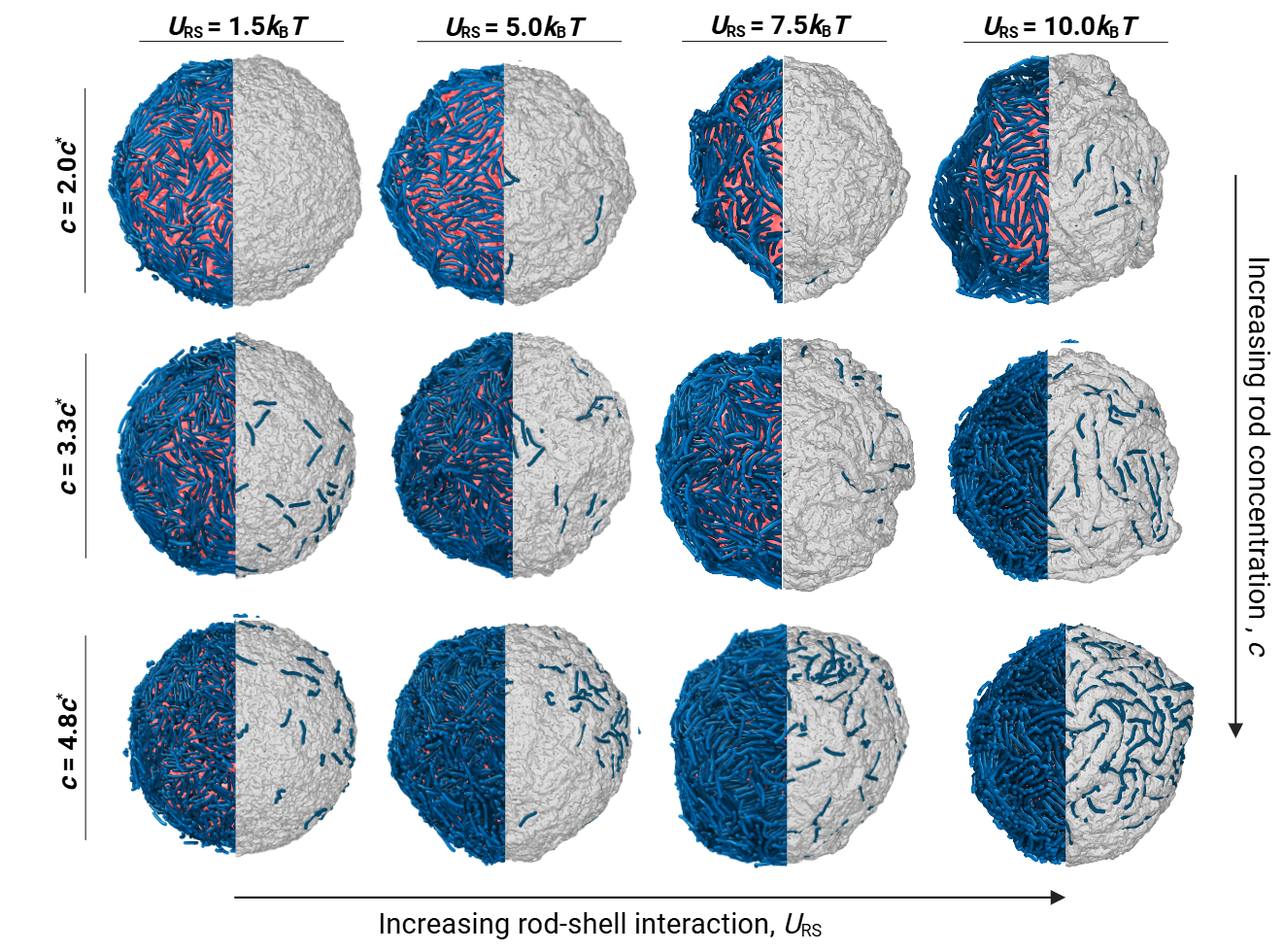}
  \caption{Exterior simulation snapshots, shell removed (right) and shell exterior(left), for various rod concentrations and $U_\mathrm{RS}$. Shape distortions increase with increasing $U_\mathrm{RS}$ and above the overlap concentration, $c > c^*$. }
\end{figure}

\begin{figure}[h!]
  \centering
  \includegraphics[width=\textwidth]{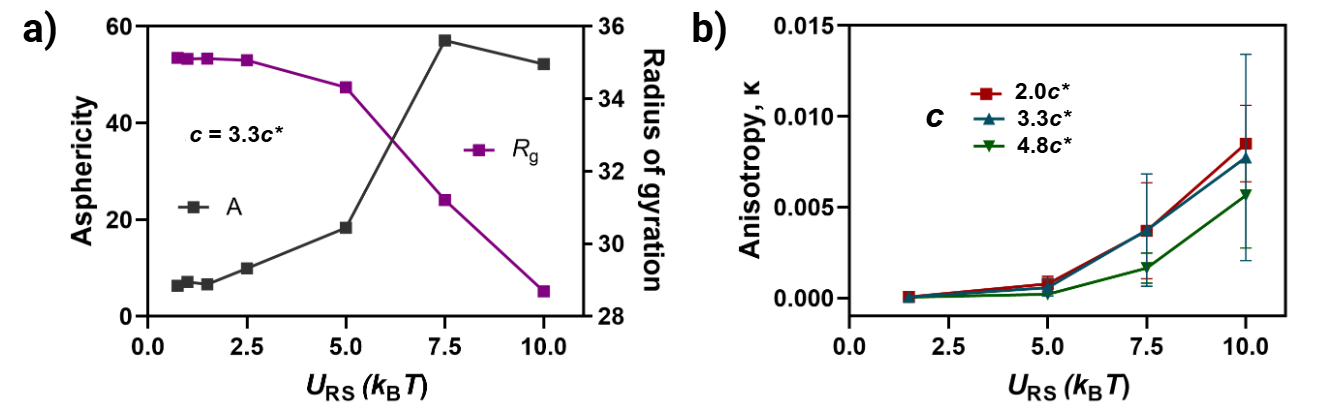}
  \caption{\textbf{a)} Radius of gyration decreases and asphericity increases with increasing rod-shell interaction strength, $U_\mathrm{RS}$, at high rod concentration, $c \approx 0.1\sigma^{-3}$. \textbf{b)} Shape anisotropy increases as a function of $U_\mathrm{RS}$ with increasing chain concentration. }
\end{figure}

\begin{figure}[h!]
  \centering
  \includegraphics[width=0.8\textwidth]{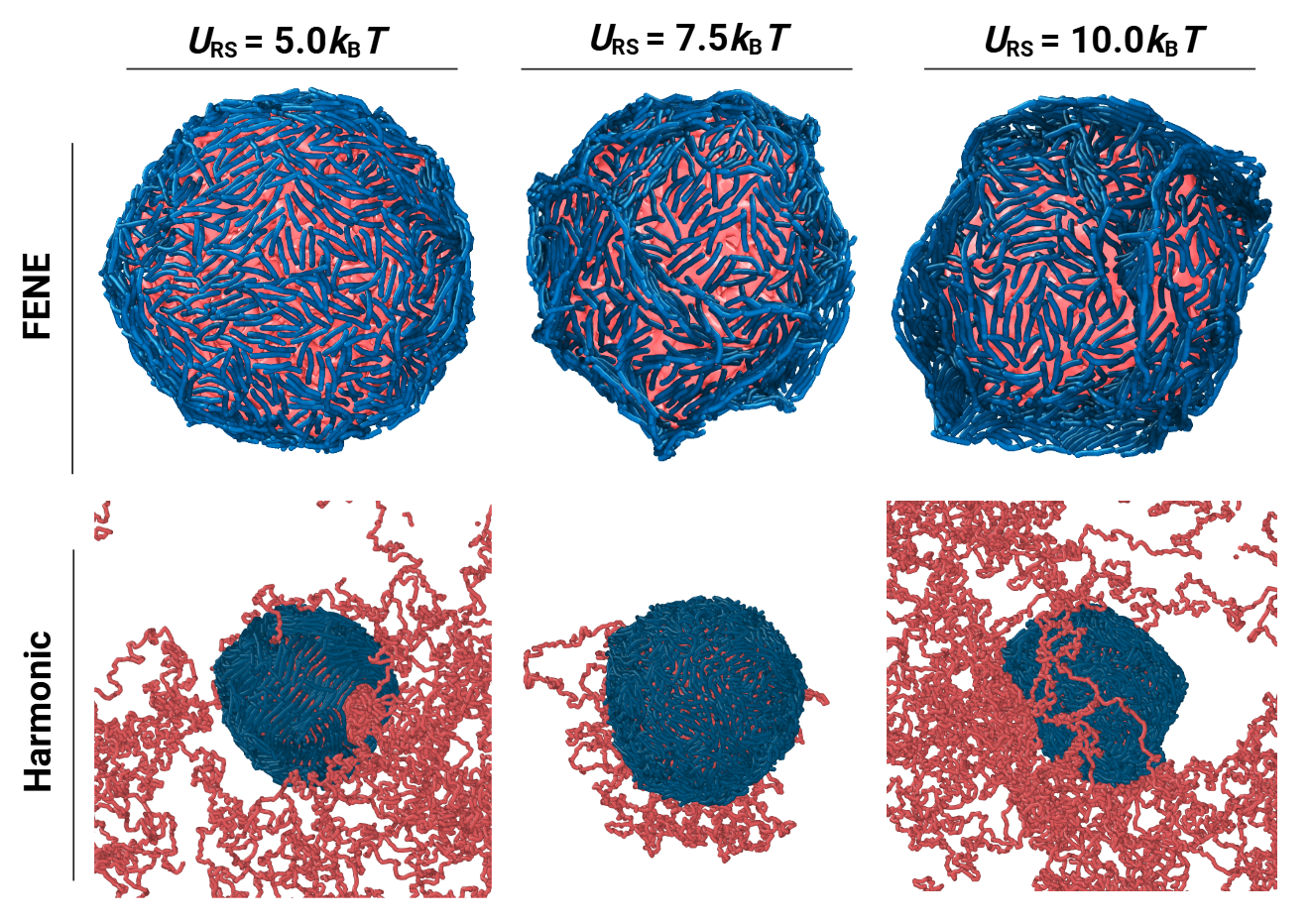}
  \caption{The elastic shell collapses inwards and polymer escapes under strong localization ($U_\mathrm{RS} = 5.0, 7.5$ and $10.0k_\mathrm{B}T$) when shell bonds are switched from FENE to extensible harmonic ones. }
\end{figure}

\begin{figure}[h!]
  \centering
  \includegraphics[width=\textwidth]{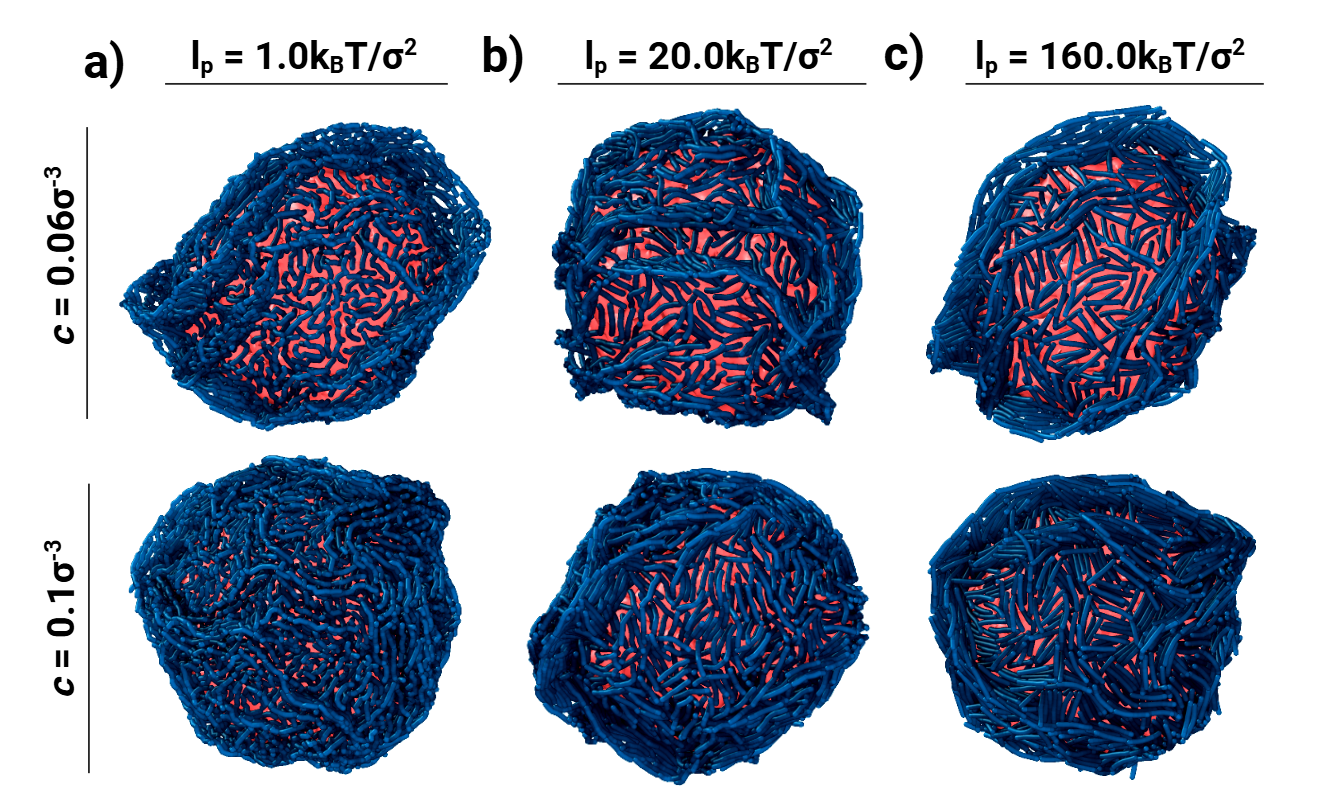}
  \caption{Distorted shapes under strong localization ($U_\mathrm{RS} = 10.0k_\mathrm{B}T$) are independent of the persistence of the chain, $l_\mathrm{p}$ at high concentrations, $c>c^*$.}
\end{figure}

\begin{figure}[h!]
  \centering
  \includegraphics[width=\textwidth]{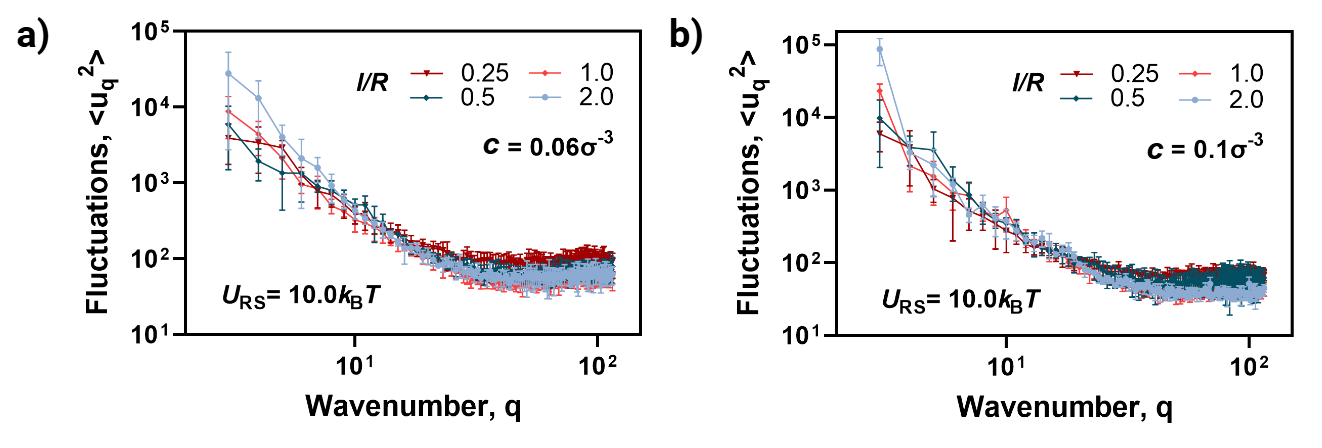}
  \caption{Fluctuation amplitudes do not vary significantly with increasing chain length, $l/R$, at high concentrations, $c>c^*$. Rod-shell attraction is also fixed to high, $U_\mathrm{RS} = 10.0k_\mathrm{B}T$.}
\end{figure}

\begin{figure}[h!]
  \centering
  \includegraphics[width=0.6\textwidth]{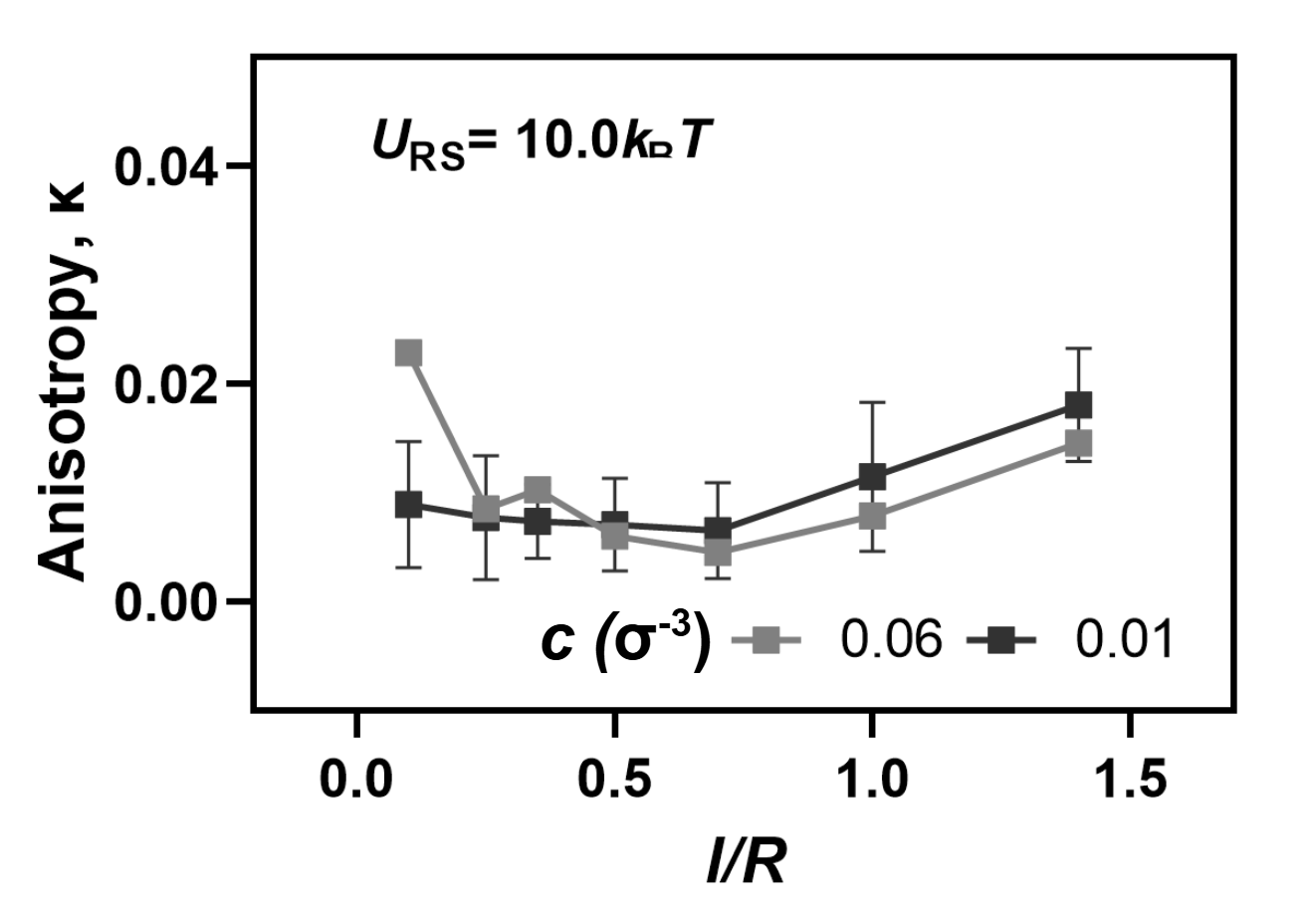}
  \caption{Anisotropy remains largely constant with increasing $l/R$ under strong localization, $U_\mathrm{RS} = 10.0k_\mathrm{B}T$. }
\end{figure}

\begin{figure}[h!]
  \centering
  \includegraphics[width=\textwidth]{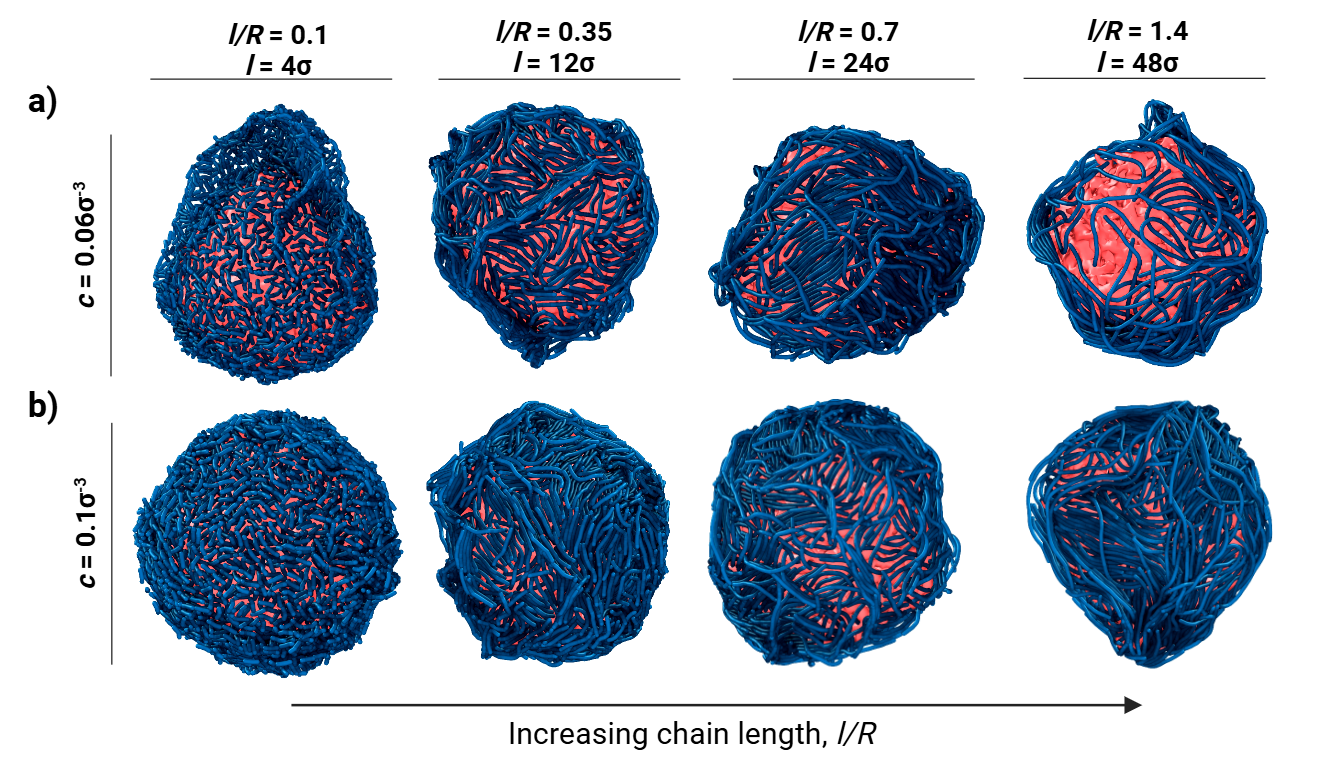}
  \caption{Simulation snapshots for distorted shell shapes at $l/R$ = 0.1, 0.35, 0.7, and 1.4 under strong localization, $U_\mathrm{RS} = 10.0k_\mathrm{B}T$. High concentrations ($c > c^*$) are chosen: $c\approx$ \textbf{a)} 0.06 and \textbf{b)} 0.1$\sigma^{-3}$. }
\end{figure}

\begin{figure}[h!]
  \centering
  \includegraphics[width=0.8\textwidth]{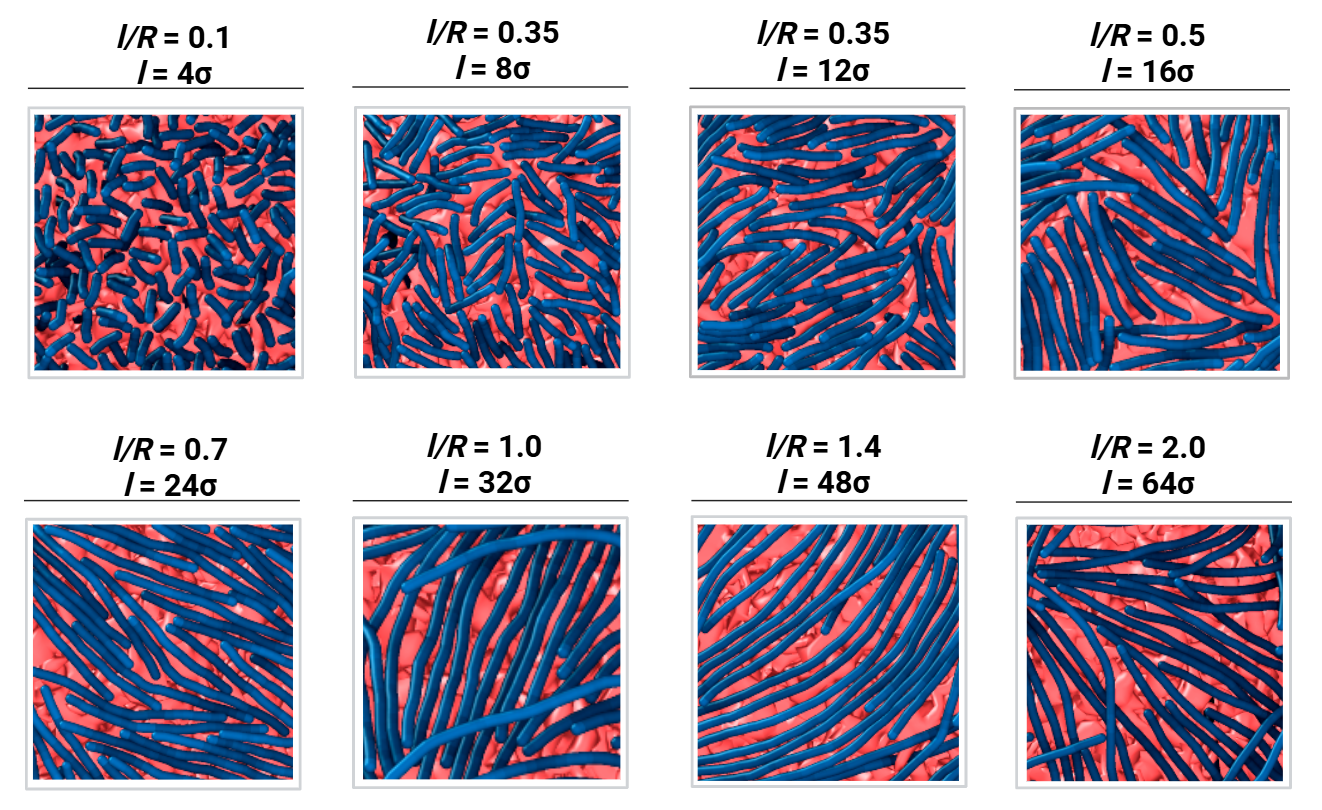}
  \caption{Organized domains emerge on the surface with increasing contour length-to-radius ratio, $l/R$, under weak localization, $U_\mathrm{RS} = 1.5k_\mathrm{B}T$, even at lower absolute concentrations, $c \approx$ 0.06$\sigma^{-3}$. This also corresponds to $c>c^*$.}
\end{figure}

\begin{figure}[h!]
  \centering
  \includegraphics[width=\textwidth]{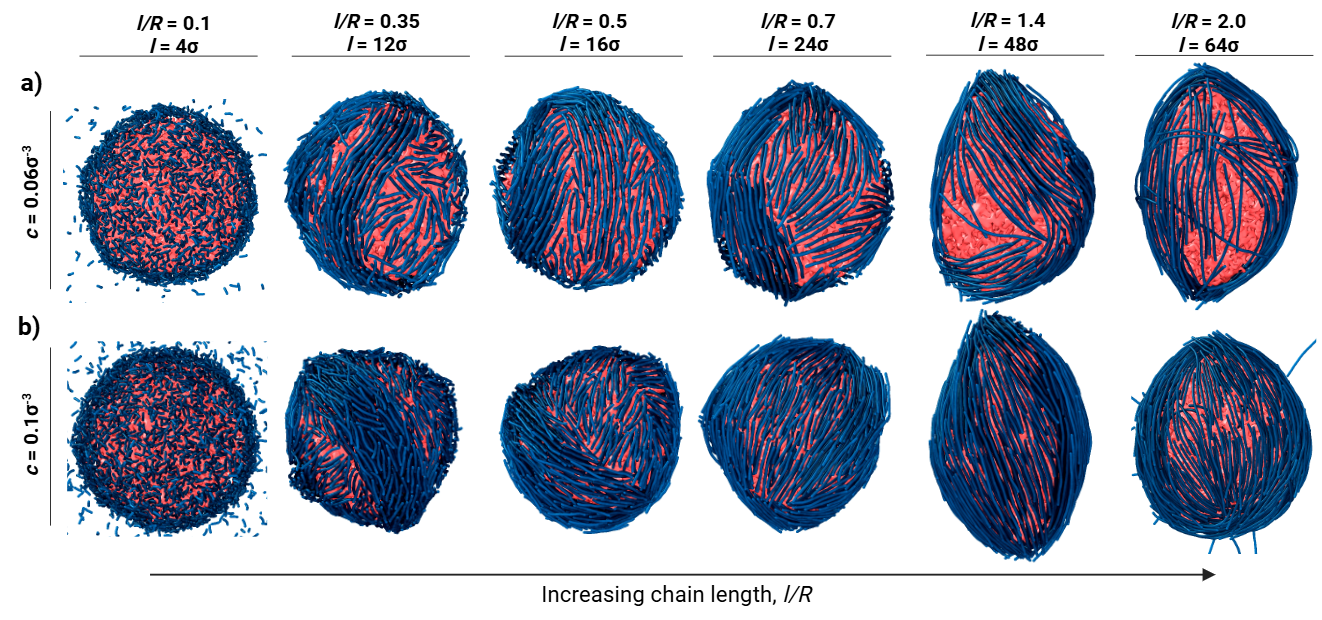}
  \caption{Shell elongates with increasing $l/R$ under weak localization when shell bonds are switched from FENE to harmonic, irrespective of chain concentration (i.e., $c \approx$ \textbf{a)} 0.06 and \textbf{b)} 0.1$\sigma^{-3}$).}
\end{figure}


\clearpage
\bibliography{supplementary} 
\bibliographystyle{supplementary}